# Systematically Deriving Domain-Specific Transformation Languages

Katrin Hölldobler
Software Engineering
RWTH Aachen University, Germany
http://www.se-rwth.de/

Bernhard Rumpe
Software Engineering
RWTH Aachen University, Germany
http://www.se-rwth.de/

Ingo Weisemöller
Software Engineering
RWTH Aachen University, Germany
http://www.se-rwth.de/

*Abstract*—Model transformations are helpful to evolve, refactor, refine and maintain models. While domain-specific languages are normally intuitive for modelers, common model transformation approaches (regardless of whether they transform graphical or textual models) are based on the modeling language's abstract syntax requiring the modeler to learn the internal representation of the model to describe transformations. This paper presents a process that allows to systematically derive a textual domain-specific transformation language from the grammar of a given textual modeling language. As example, we apply this systematic derivation to UML class diagrams to obtain a domain-specific transformation language called CDTrans. CDTrans incorporates the concrete syntax of class diagrams which is already familiar to the modeler and extends it with a few transformation operators. To demonstrate the usefulness of the derived transformation language, we describe several refactoring transformations.

## I. Introduction

Domain-specific languages (DSLs) are used in model-driven software engineering for several reasons including less complexity of the models and easier communication with domain experts [1], [2]. Nevertheless, using such models in a model-based software development process is accompanied by the need to transform these models. Therefore, a great variety of model transformation languages has become available over time realizing the various types of transformations such as refactoring, migration or translation [3], [4].

Whereas introducing DSLs for modeling is common practice, specific transformation languages are still rare. Instead, general purpose languages like Java or generic transformation languages such as ATL [5] are used. Even though there is some research done in the area of specific transformation languages [6], [7], [8], [9], [10] (also called transformations in concrete syntax) and specific languages has been developed for some domains such as building models [11] or activity diagrams [12], there is still a great need for domain-specific transformation languages (DSTLs).

Transformations in generic languages are usually difficult to write and to understand for an expert in a modeling language, because the transformations address the abstract syntax of the transformed models. Thus, they reflect the representation of the models in a modeling tool rather than the concrete syntax of the models, a representation well-suited for human users. However, developing a DSTL for a DSL requires a similar effort as developing the DSL in the first place.

In graph transformation approaches a transformation is split in two parts, a left-hand side describing a pattern and a right-hand side describing the same model part after the transformation has been applied. As described in [7] this overhead of describing unchanged model parts twice can be reduced by using an integrated notation.

Based on the idea of transformations in concrete syntax, we developed a derivation process that allows to systematically derive a DSTL for a textual modeling language defined by a grammar to reduce the effort of creating DSTLs. Similar to the derivation of delta languages described in [13], this process involves a common base language for transformation languages and derivation rules to obtain the DSTL. By applying these rules a DSTL called CDTrans was developed that is able to describe transformations for class diagrams modeled using the UML/P class diagram language [14], [15] in a problem-oriented way. Furthermore, for applying transformations modeled using the CDTrans language a generator was developed that translates those transformations to executable Java transformations.

This contribution presents the systematic derivation process of DSTLs, its application to the UML/P class diagram language and the resulting DSTL CDTrans. In order to substantiate the usefulness of the derived DSTL and to demonstrate their usage two well-known refactoring transformations taken from [16] are described using the CDTrans language: Pulling up a common attribute of subclasses to the super class and encapsulating attributes by using get and set methods. Furthermore, the application of CDTrans transformations is explained.

The remainder of this paper is outlined as follows: Sect. II briefly describes the MontiCore Language Workbench; the case examples to demonstrate the transformation language are explained in Sect. III; the transformation language itself is explained in Sect. IV, while in Sect. V the transformation language is used to solve the case examples. Sect. VI describes the systematic derivation of a transformation language and in Sect. VII the application of a CDTrans transformation is described. Subsequently, Sect. VIII discusses related work and Sect. IX concludes this paper.

K. Hölldobler is supported by the DFG GK/1298 AlgoSyn.

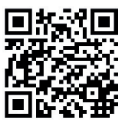
[HRW15] K. Hölldobler, B. Rumpe, I. Weisemöller:
Systematically Deriving Domain-Specific Transformation Languages.
In: Conference on Model Driven Engineering Languages and Systems (MODELS),
pp. 136-145, Ottawa, Canada, ACM New York/IEEE Computer Society, 2015
www.se-rwth.de/publications

## II. MONTICORE LANGUAGE WORKBENCH

The CDTrans transformation language as well as the generator translating transformations defined using CDTrans to executable Java transformations are implemented using the MontiCore language workbench [17]. Furthermore, the UML/P class diagram language, that the CDTrans language is designed for and the systematic derivation of the CDTrans is based on (cf. Sect. VI), is defined using a MontiCore grammar. Thus, in this section, a brief overview of MontiCore is given that explains its relevant features.

MontiCore supports the specification of textual languages by providing an EBNF like grammar format. This grammar format allows the specification of the concrete and abstract syntax of a language by a single grammar. Amongst other things, based on this grammar MontiCore generates a parser and lexer as well as language processing artifacts such as visitors and infrastructure to check context conditions [18], [19].

```
grammar CD extends Common {
  Definition =
    "classdiagram" Name "{"
      CDElement*
    "}";

  interface CDElement;

  Class implements CDElement =
    ["abstract"]? "class" Name ... ;

  Association implements CDElement =
    ... ;

  Interface implements CDElement =
    ... ;

  ...
}
```

Listing 1. Simplified Excerpt of the UML/P Class Diagram Grammar.

An excerpt of the grammar that defines the language used for modeling the UML/P [15] class diagrams in Lst. 2, 3 and 4 is shown in Lst. 1.

A MontiCore grammar consists of the keyword `grammar`, the grammar's name and surrounded by curly brackets a set of productions that define the elements of the language. In the grammar excerpt in Lst. 1 there are five productions: `Definition` (l. 2), `Class` (l. 9), `Association` (l. 12), `Interface` (l. 15) and `CDElement` (l. 7).

The class diagram itself is defined by `Definition`, whereas `Class`, `Association` and `Interface` define the classes, interfaces and associations of a class diagram. `CDElement` is a special case, i.e. an interface nonterminal that is explained later in this section.

In a MontiCore grammar as in EBNF a production is composed of a nonterminal and its definition, i.e. the right-hand side (RHS) of the production that defines the attributes and compositions within the abstract syntax tree. The RHS consists of terminals surrounded by quotation marks and nonterminals. To distinguish multiple occurrences of a nonterminal a nonterminal can be preceded with an identifier(`a:B`). For repetition the nonterminal can be marked with a star (`A*`) for arbitrary many occurrences or plus (`A+`) for at least one occurrence. Alternatives are separated by a pipe (`A|B`) and optionality is expresses by an appended question mark (`A?`).

Furthermore, MontiCore supports modularity concepts for language reuse such as language inheritance and composition (not shown here) [19], [15]. For language inheritance a grammar can extend one or more existing grammars by using the keyword `extends` followed by a comma separated list of grammars after the grammar name (l. 1). As a consequence the grammar inherits all productions of those grammars and can define further productions itself. Thus, language developers can reuse and combine existing languages and only need to define new parts of a language.

A further feature to support reuse in MontiCore grammars are interface nonterminals (l. 7). Such a nonterminal can be used like any other nonterminal within the grammar (l. 4) and is introduced by the keyword `interface` (l. 7). Interface nonterminals are an extension of alternatives, i.e defining the interface nonterminal `CDElement` is equivalent to defining a production `CDElement = Class | Association | ...`, that contains one alternative for every production that implements the interface nonterminal. Both, language inheritance and interface nonterminals are designed following the principle of object-oriented inheritance and, thus, ease the reuse and extension of existing languages [20].

## III. CASE EXAMPLE

The class diagram transformation language described in this paper is designed to describe transformations for models of the textual variant of the UML/P class diagram language. The UML/P class diagram language was developed in [14] and textually implemented in [15].

```
classdiagram ProfileManagement {

  abstract class Profile {
    public String profileName;
  }

  class Person extends Profile {
    public String name;
    public String address;
  }

  class Group extends Profile {
    public String purpose;
    public String address;
  }
}
```

Listing 2. Running Example.

The diagram that serves as the running example is shown in Lst. 2. It describes a very simple profile management e.g. for a social network. The class diagram comprises an abstract class `Profile` and two normal classes, `Person` and `Group`, both of which extend the abstract class `Profile`. Profiles have profile names, a person has a name and an address and a group has a purpose and also an address.

To demonstrate the transformation language CDTrans two well known example transformations from [16] have been chosen: Pulling up a common attribute and encapsulation of attributes. These examples have been chosen as they need to match specific parts of the model as well as modify, create, relocate and delete model parts.

*A. Case 1: Pulling Up Attributes*

Pulling up an attribute is a refactoring in which two subclasses of a class share a common attribute. The rule described in [16] (here called Pull Up Field) specifies that the common attribute is removed from both subclasses and relocated in the super class. In the running example both classes `Group` and `Person` have the attribute `address` that after applying this rule is located in their super class `Profile` as depicted in Lst. 3.

```
classdiagram ProfileManagement {                           CD

  abstract class Profile {
    public String profileName;
    public String address;
  }

  class Person extends Profile {
    public String name;
  }

  class Group extends Profile {
    public String purpose;
  }
}
```
Listing 3. Class Diagram after Applying the Pull Up Transformation.

Certainly, this rule can be extended such that an arbitrary number of subclasses can exist and in case all have the same attribute it is pulled up as for example done in [21], but within the scope of this case study we limit this case to two subclasses but in addition require that no further subclass must exist.

*B. Case 2: Encapsulation of Attributes*

```
classdiagram ProfileManagement {                           CD
  ...
  class Group extends Profile {
    private String purpose;
    private String address;

    public String getPurpose();
    public void setPurpose(String purpose);
    public String getAddress();
    public void setAddress(String address);
  }
}
```
Listing 4. Class Diagram after Applying the Encapsulating Transformation.

The second case also taken from [16] is about encapsulating attributes, i.e. the visibility of public attributes is changed to private and the public access is established by adding public get and set methods for the attributes. Regarding the running example an excerpt of the resulting class diagram is given in Lst. 4. Due to space limitations the effect is shown for the class `Group` only[1].

---
[1]Please note that we have limited this example to the method signature here, even though the method bodies could also be added by a transformation.

## IV. CDTRANS

Complex transformations usually consist of transformation rules that are applied using some kind of control structures or application strategies [4], [3]. The CDTrans transformation language is a language that allows textually describing transformation rules for UML/P class diagrams in a class diagram specific way. The application of these rules is dedicated to the user and further explained in Sect. VII.

CDTrans realizes a graph-based transformation approach. It allows to describe endogenous, in-place model transformations. In those approaches, transformation rules consists of a left-hand side (LHS), i.e. the pattern to be found in the model and a right-hand side (RHS), i.e. the same model part after the transformation is applied (see [22], [23]).

Following the principle described in [7], [24], the CDTrans is based on the concrete syntax of the class diagram language instead of its abstract syntax and uses an integrated notation of the LHS and RHS of a transformation rule. In the following the syntax to describe patterns, modifications, constraints and negative application conditions is explained.

*A. Pattern*

As the notation of transformations is based on the concrete syntax of the class diagram language a pattern without modifications is quite similar to the model part being described. For the first case example the three classes as well as the common attribute `String address` of two classes need to be matched. A pattern description for this part of the model is given in Lst. 5. Even though the same order as in the model was chosen for the classes, this is not necessary for the matching process.

```
class Profile;                                             MTR

class Person extends Profile {
  public String address;
}

class Group extends Profile{
  public String address;
}
```
Listing 5. Pattern for the Classes `Person` and `Group` that Extend the Class `Profile`.

The pattern requires those three classes, their relation, i.e. `Person` and `Group` extend the class `Profile`, and the common attribute. As can be seen from this example there is no need to include irrelevant details in the pattern. Additional attributes as well as the keyword abstract that are present in the model have been omitted in the pattern. Those details are not relevant for the model part that should be matched. Omitting those model parts is not equivalent to requiring a non-abstract class or not allowing the classes to have any further attributes. Instead, there is nothing said about whether the classes are abstract or have any further attributes. Thus, arbitrary classes fulfilling the described properties are possible matches. Furthermore, there is no need to start at the class diagram level, e.g. this pattern starts at the class level.

It describes the correct part of exactly this model but in general a transformation rule needs to be more general and should be applicable to different models. In order to allow more general transformation rules, the CDTrans transformation language has a concept called schema variables. Those variables can be used to (1) bind a model element to a variable which will be useful e.g. for modifications such as moving elements and (2) to abstract from concrete values such as names.

```
1  class $parent;                                          MTR
2
3  class $_ extends $parent {
4    Attribute $A1;
5  }
6
7  class $_ extends $parent {
8    Attribute $A2;
9  }
10
11 where {
12   $A1.deepEquals($A2)
13 }
```

Listing 6. Pattern with Schema Variables and Constraint.

Lst. 6 is a generalization of the pattern given in Lst. 5. Instead of the concrete names of the classes schema variables `$_` and `$parent` are used for the names of the classes and the attributes are replaced by the schema variables `$A1` resp. `$A2` plus its type `Attribute` (which is discussed in Sect. VI-E).

A schema variable is composed of a `$`-sign and an arbitrary name for this variable. Schema variables can be used for all elements of a class diagram, e.g. classes, attributes, methods or names. The CDTrans language offers the following two options to bind variables:

$$ElementType\ SchemaVariable; \qquad (1)$$
$$ElementType\ SchemaVariable\ [[\ Element\ ]] \qquad (2)$$

The first variant can be understood as a black box variant, an arbitrary element of the corresponding type will be bound to the variable during the pattern matching. This variant is used for the attributes in Lst. 6. The second variant instead is a white box variant. In this case the concrete syntax of the element is given in the double square brackets. For demonstrating this variant it was used for the attribute in Lst. 8.

Furthermore, the CDTrans language has a special treatment of schema variables for names. On the one hand schema variables for names only consist of the variable itself without the type; on the other hand using the same variable for two name occurrences in a model requires equality of the two names but not the same identity as it is the case for other two occurrences of the same schema variable. This variant is used in Lst. 6 (`$parent`) and Lst. 8 (`$type` and `$attrname`).

The `$_`-variable is a special case, it just replaces the concrete value but is does not bind any value. Thus, the pattern abstracts from the concrete names, but does not enforce both classes to have the same name.

### B. Negative Elements

Another feature of the CDTrans language are negative elements. Negative elements are negative application conditions [25], [26]. Negative elements are elements that must not be present at the specified position of the model to find a match for a pattern. Thus, a transformation can only be successfully applied if there is no match for the negative elements described in a transformation. The CDTrans language provides the following syntax to describe negative elements:

$$\texttt{not}\quad [[\ Element\ ]]$$

The requirement of the "Pulling-Up-Attributes" case (i.e. there exists no further subclass) is a typical use case for negative elements. The positive variant would require to describe all other available classes or to formulate a rather complicated constraint to achieve the same effect. For the case example the following line needs to be added to complete the pattern:

$$\texttt{not [[ class \$\_ extends \$parent; ]]}$$

### C. Modifications

The CDTrans language uses an integrated notation for the LHS and the RHS part of a transformation. To describe modifications the CDTrans offers the following operator:

$$[[\ Element?\ :\text{-}\ Element?\ ]]$$

This operator looks quite similar to a description of the LHS/RHS form and has a similar meaning. The element left of the `:-` is replaced by the element right of it. But in contrast to the common LHS/RHS form, this operator can be used for elements instead of complete patterns. Furthermore, this operator can be used to create or delete elements. In the former case the left-hand side is empty, in the latter the right-hand side is empty. The replacement of keywords and names is done by the same operator. Examples for replacements are shown in line 2 and 6 of Lst. 7. Here, the attribute `$A1` is moved from one class (l. 6) to another (l. 2). In line 4 of Lst. 8 the keyword `private` is deleted and `public` is added.

### D. Application Constraints and Assignments

The CDTrans language allows to specify a where-block that allows to define an application constraint for the transformation and assign values to schema variables. If an application constraint is defined, a transformation is only applied if (1) the pattern is found in the model and (2) the application constraint holds. The syntax of a where-block is the following:

$$\texttt{where}\quad \{\ Assignment * BooleanExpression?\ \}$$

A where-block consists of the keyword `where` followed by an arbitrary number of assignments of schema variables and an optional boolean expression (i.e. the application constraint) surrounded by curly brackets.

The assignments allow to assign values to schema variables that are not assigned during pattern matching, i.e. that are only part of the RHS of a modification (Sect. IV-C). An example for an assignment is shown in Line 12-13 of Lst. 8 that shows a solution for the case example 2. Here, the assignments concatenate the name of the get and set methods.

Within the boolean expression all elements bound to schema variables can be used to formulate an application constraint. An example of an application constraint is given in Line 15 of Lst. 7 that shows a solution for the first case example. This constraint requires the two attributes of the subclasses to be equivalent. As can be seen from this example within the constraint the signature of the abstract syntax of the model elements can be used. Furthermore, as CDTrans transformations are translated to Java any static helper methods can be used.

## V. SOLUTIONS FOR THE CASE EXAMPLES

In this section possible realizations of the two case example transformations using the CDTrans transformation language are explained.

### A. Case 1: Pulling Up Attributes

Lst. 7 shows a transformation realizing the first case example transformation, i.e. pulling up a common attribute.

```
class $parent {
  [[ :- Attribute $A1; ]]
}

class $_ extends $parent {
  [[ Attribute $A1; :- ]]
}

class $_ extends $parent {
  [[ Attribute $A2; :- ]]
}

not [[ class $_ extends $parent; ]]

where { $A1.deepEquals($A2) }
```

Listing 7. Transformation to Pull up a Common Attribute.

Line 5-7 and 9-11 describe a pattern of two classes with arbitrary names (indicated by the variable $\_) that extend the same super class with an arbitrary name that is bound to the variable $parent (same variables for names means equality). Both classes furthermore have an attribute bound to the variable $A1 resp. $A2 (choosing the same variable for the attributes would require identity). The constraint in line 15 requires this two attributes to be equal. The pattern of the super class is described in line 1. The modification of this transformation is described by the lines 2, 6 and 10 where the replacement operator is used to remove the attributes $A1 and $A2 from the subclasses (l. 6 and 10) and add the attribute $A1 to the super class (l. 2). Finally, the negative element in line 13 enforces that the transformation is only applicable if there is no further class that extends the super class $parent.

### B. Case 2: Encapsulation of Attributes

Lst. 8 shows a transformation realizing the second case example transformation, i.e. encapsulating attributes.

Lines 1-9 describe the pattern of a class with an arbitrary name (indicated by the $\_ variable) with a public attribute whose type and name are arbitrary but bound to the variables $type resp. $attrname. For demonstrating a white box schema variable the attribute is bound to the variable $A (l. 3-5). The modification in line 4 changes the public visibility of the attribute to private. In line 7 the method signature for the get method is added and in line 8 the signature of the set method is added. For the return type of the method as well as for the type of the attribute of the set method the type of the attribute is used. The attribute name is used for the parameter name. The transformation uses variables for the names of the methods that are not defined in the pattern part of the transformation ($set and $get). The assignments within the where-block are used to assign the values (here Strings for the names of the methods) to those variables. The names for the methods are composed of the prefix get resp. set and the name of the attribute. Please note that this will lead to getaddress and setaddress instead of camel cased names but this can easily be corrected by using a Java class that capitalizes the attribute name within the String concatenation, e.g. WordUtils.capitalize(...) of the Apache Commons [27] or the StringTransformations.capitalize(...) of MontiCore.

```
class $_ {

  Attribute $A [[
    [[public :- ]] [[ :- private]] $type $attrname;
  ]]

  [[ :- public $type $get ();]]
  [[ :- public void $set ($type $attrname);]]
}

where {
  $get = ("get").concat($attrname);
  $set = ("set").concat($attrname);
}
```

Listing 8. Transformation to Encapsulate Attributes

## VI. SYSTEMATIC DERIVATION OF A DSTL

In this section the derivation rules for systematically creating DSTLs and the common grammar for DSTLs are described. To ease the development of DSTLs a grammar called `TFCommons` has been developed that defines the modeling language independent parts of a derived DSTL. This grammar is used as a base grammar for DSTLs derived according to the derivation rules described later in this section. The derivation can be automated by e.g. traversing the modeling language grammar using a visitor and thereby creating the transformation language grammar according to the derivation rules.

### A. Common Base Grammar for Transformation Languages

Lst. 9 shows an excerpt of the TFCommons grammar that defines the modeling language independent part of a DSTL. This grammar basically provides the syntax for schema variables, names and their replacements, and the where-block that allows defining an application constraint and assignments for schema variables.

The `TFIdentifier` defined in line 3-4 is used within the transformation language for names (cf. derivation rule 4a). The nonterminal `Ident` is either a schema variable or a concrete name (to be matched in the model). Thus, the `TFIdentifier` provides the replacement operator for names (concrete ones or those bound to schema variables). Please note that names can only be replaced but not deleted or created.

Furthermore, the nonterminal `Where` provides the syntax to describe the where-block of a transformation. It allows

```
1  grammar TFCommons extends Common {
2
3   TfIdentifier =
4    Ident | "[[" Ident ":-" rhs:Ident "]]";
5
6   Where = "where" "{"
7             Assignment*
8             constraint:BooleanExpression?
9           "}";
10
11  Assignment = SchemaVar "=" value:Expression ";";
12  ...
13 }
```

Listing 9. Simplified Excerpt of the Common Grammar for DSTLs.

several assignments of schema variables (cf. l. 7 and 11) and an application constraint (cf. l. 8).

*B. Derivation Rules*

The grammar for a DSTL is created by systematically applying derivation rules. It consists of a nonterminal for the transformation rule itself and nonterminals to describe pattern, modifications and negative elements of all nonterminals and optional keywords (i.e. keywords that can be set, removed or replaced) of the modeling language. These derivation rules are explained in the following sections.

*Interface Nonterminals:* As the initial step, for every nonterminals and every optional keyword such as `abstract` or `public` an interface nonterminal is created. This nonterminal is used to bundle the nonterminals created in further derivation rules. Thus, the first rule is split into three subrules:

**1a.** *For every interface nonterminal $N \in L$ create an interface nonterminal $N$ in the transformation language $TL$. If $N \in L$ extends interface nonterminals $I \in L$, $N \in TL$ extends $I \in TL$.*

**1b.** *For every nonterminal $N \in L$ create an interface nonterminal $N$ in the transformation language $TL$. If $N \in L$ implements interface nonterminals $I \in L$, $N \in TL$ extends $I \in TL$.*

Using the same name for interface nonterminals of $TL$ as for the nonterminals in $L$ was a deliberate decision to ease the derivation rule for the nonterminals describing the pattern (cf. rule 4a). Furthermore, the relation between normal and interface nonterminals, i.e. nonterminals implementing interface nonterminals and interface nonterminals extending other interface nonterminals is reflected by the interface nonterminals created for the transformation language.

**1c.** *For every optional keyword $k \in L$ create an interface nonterminal $K$ in the transformation language.*

The name of this interface nonterminal is derived from the keyword the same way as MontiCore derives the name for attributes corresponding to optional keywords but capitalized, e.g. `"abstract"` becomes `Abstract`, `"+"` becomes `Plus`, and `private:"-"` becomes `Private`.

*Modification:* In order to allow modification of every model element nonterminals providing the replacement operator are created for every nonterminal and every optional keyword of the modeling language $L$. Modification for names are provided by the common grammar explained in Sect. VI-A. The replacement for keywords and nonterminals differ, thus, the second derivation rule is split into the following two subrules:

**2a.** *For every (interface) nonterminal $N \in L$ create a nonterminal `N_Rep` of the following form:*

```
N_Rep implements N =
    "[[" lhs:N? ":-" rhs:N? "]]";
```

By introducing the nonterminals `N_Rep` for nonterminals of $L$ a transformation can describe replacing one element by another one, adding an element (LHS is empty) and removing an element (RHS is empty).

**2b.** *For every optional keyword $k \in L$ create a nonterminal `K_Rep` of the following form:*

```
K_Rep implements K =
"[[" "k" ":-" "]]" | "[[" ":-" "k" "]]";
```

This rule handles all optional keywords separately, thus, the replacement of keywords differ from the replacement for nonterminals as keywords can either be added or removed but not be replaced as replacing a keyword by itself is not needed. Please note that for some languages this systematic handling of keyword might be counterintuitive e.g. adding private and removing public could be more intuitively modeled by replacing public by private. Therefore, depending on the modeling language a manual adaptation of the systematically derived language could further improve the DSTL.

*Negative Elements:* For every nonterminals and every optional keyword, nonterminals to describe negative elements are created. The following two subrules describe the structure that is created for those nonterminals:

**3a.** *For every (interface) nonterminal $N \in L$ create a nonterminal `N_Neg` of the following form:*

```
N_Neg implements N = "not" "[[" N "]]";
```

**3b.** *For every optional keyword $k \in L$ create a nonterminal `K_Neg` of the following form:*

```
K_Neg implements K = "not" "[[" K "]]";
```

*Pattern:* Finally, the concrete syntax of the modeling language needs to be transferred to the transformation language to describe pattern. Thus, nonterminals for the pattern are created according to the two subrules:

**4a.** *For every nonterminal $N \in L$ create a nonterminal `N_Pattern` of the following form:*

`N_Pattern` **implements** `N =`
$SyntaxOfN$ |
`"N"` `SchemaVar` `(";"` | `"[["` $SyntaxOfN$ `"]]");`

*where $SyntaxOfN$ is a modified copy of the productions RHS of the nonterminal `N`. Within this copy all occurrences of*

keywords are replaced by the interface nonterminal introduced for the keyword and all name occurrences are replaced by the nonterminal `TFIdentifier` defined in the `TFCommons` grammar.

Please note that nonterminals occurring in the copy need not be changed as they refer to the interface nonterminals of the transformation language created by rule 1. Thus, they already allows patterns, modification or negative elements.

**4b.** *For every interface nonterminal $N \in L$ create a nonterminal `N_Pattern` of the following form:*

```
N_Pattern implements N =
            "N" SchemaVar ";";
```

Interface nonterminals define no concrete syntax itself, instead nonterminals implementing those nonterminals define the concrete syntax. However, this rule provides the black box variant of schema variables for interface nonterminals. By providing this variant a transformation is able to match an arbitrary element that implements the interface nonterminal in $L$. An example use case is an interface nonterminal `Type` that is implemented by nonterminals representing different kinds of types. If the concrete type is irrelevant a transformation can match the `Type` element itself.

**4c.** *For every optional keyword $k \in L$ create a nonterminal `K_Pattern` of the following form:*

```
K_Pattern implements K = "k";
```

As the concrete syntax of a keyword is the keyword itself, there is no need for schema variables for keywords. Thus, the nonterminal `K_Pattern` has no alternatives.

*Transformation Rule:* As the last step a nonterminal that represents a complete transformation rule is needed. Thus, this rule creates a nonterminal that combines all nonterminals to form a transformation rule:

**5.** *Create a nonterminal `TFRule` of the following form:*

```
TFRule = (AlternativeOfNTs)* Where?;
```

where $AlternativeOfNTs$ is an alternative of all interface nonterminals created for nonterminals of $L$.

This nonterminal allows all elements of a model as top level elements in a transformation rule. Therefore, a transformation rule needs not start at the top level element of a model, e.g. the class diagram definition. Instead, the transformation developer can concentrate on the concrete model part that should be transformed. Furthermore, this nonterminal adds the where-block to the transformation rule.

### C. Context Condition

The systematic derivation in conjunction with the reuse of the nonterminals structure of the modeling language already achieves that the structure of a transformation is always conform to a model structure, e.g. classes cannot be specified within other classes, modifications may only add the correct type of element, etc.. However, there are some context conditions regarding the transformation specific parts of the transformation that must hold for a transformation to be valid.

*1) A schema variable must be unique.:* As a schema variable is bound during pattern matching or is assigned within the where-block of a transformation those variables must be unique. Nevertheless, they may occur within the pattern/assignment, the RHS/and the constraint referring to the same element. Exception: Two occurrences of the same schema variable for names as this means equality of the name not the same identity.

*2) A schema variable on the RHS of a modification must either occur within the pattern or be assigned within the where-block.:* By formulating this context condition it is guaranteed that a modification that has an element on its RHS actually adds or moves this element. Allowing not assigned variables on the RHS would either shadow a mistake or change the behavior of the replacement operator. Therefore, a not assigned variable must be reported as an error.

*3) A schema variable used within the where-block must exist.:* Schema variables can be used inside a where-block either as part of the application constraint or be assigned with a value. In case a schema variable is used within the constraint it must be part of the pattern. If a variable is assigned it must exist on the RHS of a modification only.

*4) There must not occur negative elements on the RHS of a modification.:* Negative elements are elements that must not be present at the specified position, thus, they can only constrain the pattern part of a transformation. For modifications creating new elements there is no need for negative elements. For modification moving elements the negative elements should be specified in conjunction with the pattern describing the elements that is moved.

*5) Negative elements must not be nested.:* To avoid double or more negation of elements and thus keep this operator simple, this context condition forbids using negative elements within negative elements. This limitation is compensated by the application constraint that allows to formulate an arbitrary boolean expression using the schema variables, their signature and arbitrary static Java methods (cf. Sect. IV-D).

*6) There must not occur a modification within a negative element.:* Negative elements are elements not to be present, hence, modifying elements that are not present in a model is not possible. Hence, modifications within negative elements should not be possible and, thus, be reported as an error.

Depending on the modeling language there might be further conditions that must hold with regard to the modeling language. An example of such a condition could be that a transformation must not remove the visibility of a method without adding a new visibility.

### D. Application of the Derivation Rules

In this section the derivation process is demonstrated by applying the derivation rules to the class diagram grammar shown in Lst. 1. Lst. 10 shows a simplified excerpt of the derived grammar defining the CDTrans language.

As shown in Lst. 10 the derived grammar extends the common grammar for DSTLs TFCommons (l. 1) providing replacements for names, schema variables and the where-block.

Applying the **first derivation rule** (1a-1c) the interface nonterminals to bundle the pattern, modification and negative element nonterminals are created (l. 7-11). According to the rule, interface nonterminals are created for the interface nonterminals (cf. CDElement in l. 7), normal nonterminals (cf. Definition, Class, l. 8-9) and optional keywords[2] (cf. Abstract, l. 10). Furthermore, the implements relation of Class and CDElement is reflected by the Class interface nonterminal extending the CDElement nonterminal.

According to the **second derivation rule** (2a-2b) the replacement nonterminals (cf. Definition_Rep, l. 14-16 and Abstract_Rep, l. 35-37) are created.

After applying the **third derivation rule** nonterminals for describing negative elements are created (cf. Definition_Neg in l. 19-20, Abstract_Neg in l. 40-41).

Nonterminals for describing patterns in concrete syntax as well as for schema variables are created according to the **fourth derivation rule** (cf. Definition_Pattern in l. 23-32, Abstract_Pattern in l. 44-45). As described by this rule, occurrences of names are replaced by TFIdentifier (cf. l. 24 and 29). References to other nonterminals need not be changed (cf. l. 25 and 30). Furthermore, occurrences of keywords are replaced by their corresponding interface nonterminals (not shown here).

Finally, according to the **fifth derivation rule** the nonterminal describing a transformation rule is created (cf. l. 4-5).

### E. Discussion

The process of deriving a DSTL is completely systematic and based on the grammar of the modeling language $L$. The main part of the DSTL is based on the concrete syntax of $L$. For typing and abstraction purposes the nonterminal names of $L$ are used for the typing of schema variables. Thus, these names become part of the concrete syntax of the DSTL. A schema variable on its own (i.e. without the type) is sufficient if the type can be inferred from the environment of the variable but for nonterminals in alternatives the type cannot be inferred correctly (at least for the black box variant). Thus, the systematic derivation includes types for schema variables. However, the derived grammar can be extended manually using the language inheritance feature of MontiCore to refine this part of the concrete syntax. Furthermore, using inheritance keywords such as not, delimiters or operators can be refined to eliminate possible conflicts regarding keywords of $L$.

The structure of the derived DSTL depends on the structure of $L$. The abstract syntax of $L$ might contain folded or expanded productions. Even though this does not affect the concrete syntax of $L$, it leads to less or more nonterminals in the DSTL as replacements, negations and variables are created for every nonterminal of $L$. Thus, if the DSTL's structure is too fine-grained or too coarse, restructuring $L$ solves this issue.

---

[2] class and classdiagram are also keywords but not optional.

```
grammar CDTrans extends TFCommons {

  //5
  TFRule = (Definition | Class | ... )*
          Where?;

  interface CDElement;            // 1a
  interface Definition;           // 1b
  interface Class extends CDElement; // 1b
  interface Abstract;             // 1c
  ...

  // 2a
  Definition_Repl implements Definition =
    "[[" lhs:CDDefinition?
    ":-" rhs:CDDefinition? "]]";

  // 3a
  Definition_Neg implements Definition =
    "not" "[[" CDDefinition "]]";;

  // 4b
  Definition_Pattern implements Definition =
    "classdiagram" TfIdentifier "{"
      CDElement*
    "}"
    | "Definition" SchemaVar ";"
    | "Definition" SchemaVar "[["
      "classdiagram" TfIdentifier "{"
        CDElement*
      "}"
    "]]";

  // 2b
  Abstract_Rep implements Abstract =
    "[[" lhs:Abstract ":-" "]]"
    | "[[" ":-" rhs:Abstract "]]";

  // 3b
  Abstract_Neg implements Abstract =
    "not" "[[" Abstract "]]";

  // 4c
  Abstract_Pattern implements Abstract =
    "abstract";
  ...
}
```

Listing 10. Simplified Excerpt of the CDTrans Grammar

Another problem are nonterminals that do not define any mandatory concrete syntax. As all nonterminals are allowed as top level elements within a transformation rule such nonterminals might result in parsing problems as the parser could infinitely often parse an empty nonterminal. However, restructuring $L$ or excluding those nonterminals from the alternative of the TFRule nonterminal solves this problem.

As MontiCore supports language inheritance, a modeling language $L$ might extend another language $SL$. Within this derivation process only nonterminals defined in $L$ are considered. However, the same derivation process can be used to derive a transformation language $TSL$ for $SL$. Thus, using language inheritance the transformation language $TL$ for $L$ can extend $TSL$ such that $TL$ inherits all nonterminals to describe transformations of elements defined by $SL$.

Finally, deriving nonterminal names from keywords might result in naming clashes. For example, a nonterminal Association that contains the keyword association would result in two nonterminals called `Association` within the transformation language grammar. In order to avoid this either the modeling language could be adapted or the derivation rules for keyword can add an additional suffix such as `Keyword`.

## VII. APPLYING A CDTRANS TRANSFORMATION

The CDTrans transformation language is accompanied with a generator that takes a transformation as input and generates a Java class that is able to execute the transformation. The generated Java class realizes the pattern matching for the pattern part of the transformation using a search plan based pattern matching approach [28], [29] and in case a match has been found applies the changes described by the transformation. This workflow from creating a transformation to applying it is depicted as an activity diagram in Fig. 1.

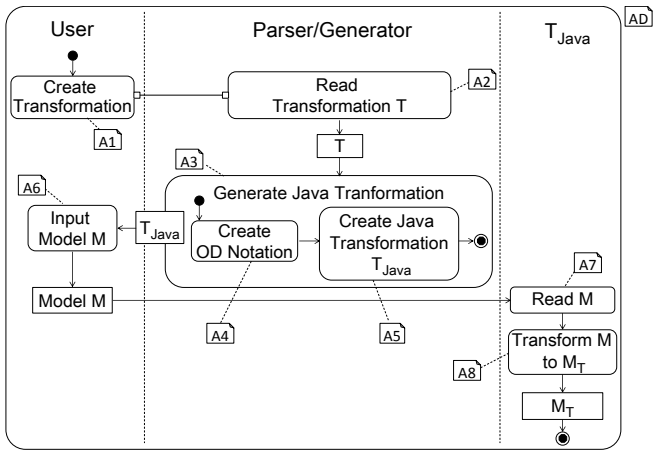

Figure 1. Workflow to Translate and Apply a Transformation.

After creating a transformation T (A1) the transformation will be read in by the generator (A2) and translated to a Java transformation $T_{Java}$ (A3). This transformation is returned to the user/modeler who provides a model M (A6) as input for the transformation $T_{Java}$. The transformation $T_{Java}$ reads in the model (A7), does the pattern matching, modifies the model and thereby produces the transformed model $M_T$ (A8).

The generator translating those transformations is implemented using MontiCore [17]. A two-step process was chosen to translate the domain-specific transformation (DST). In the first step (A4) the DST is translated to a generic object diagram based notation describing the LHS and RHS of the transformation rule. In the second phase (A5) of the generation this generic notation is used to generate the Java transformation class. This translating generator is composed of two interacting generators where the outcome of the first generator (i.e the generic LHS/RHS notation) is used as the input for the second generator. Choosing this two-step process allows a reuse of the second generator for further transformation languages similar to CDTrans and different front ends that could translate to the LHS/RHS notation and use the second generator, e.g. delta languages [13]. We forego the details of this two generators and plan to detail this approach in a further publication.

## VIII. RELATED WORK

Similar to e.g. PROGRES [30], Fujaba [31], eMoflon [32] and Henshin [33] the CDTrans transformation language is a graph transformation approach and supports endogenous model transformations [4]. But in contrast to those approaches CDTrans reuses the concrete syntax of the modeling language.

A DSTL for UML diagrams using the graphical concrete syntax of UML is presented in [8]. However, there seems not to be any implementation though. In [34] a generator framework to translate domain-specific transformations to executable Java transformations is proposed but there is no explanation how this should be achieved in practice.

In [6], [35], the metamodel of a pattern language able to describe the LHS/RHS pattern of a transformation is generated based on the metamodel of a modeling language. Compared to the systematic derivation of abstract and concrete syntax presented in this paper the transformation language developer has to create a concrete syntax for the abstract syntax.

The approach presented in [36], [9], generates a graphical transformation language based on a graphical modeling language. To reuse the concrete syntax the DSTL developer needs to link the abstract syntax to the concrete syntax of the modeling language. In [12] a DSTL for UML 2 activity diagrams [37] based on this approach is presented. Another tool for domain-specific transformations for graphical models is AToMPM [10] based on T-Core [38].

Inferring model transformations from examples [39] is a similar approach to DSTLs. Example source and target models are used as the LHS resp. RHS and are aligned by the user. However, to correct and generalize a transformation, the user must operate on the abstract syntax.

Similar to the systematic derivation presented here, is the approach presented in [13]. This approach derives delta languages from modeling languages. However, the derivation process is specific for delta languages.

The approach presented in [40] provides transformation primitives that can be combined to create a DSTLs. This approach focuses on creating transformation languages by combining and configuring transformation language elements rather then defining a systematic on how to create a language.

## IX. CONCLUSION

Model transformations are helpful to evolve, refactor, refine and maintain (domain-specific) models. While DSLs are normally intuitive for modelers, common model transformation approaches use the abstract syntax of the modeling language, that is usually hidden from the modeler, to describe transformations. To alleviate the use of abstract syntax, in this paper a process to systematically derive a DSTL was presented that looks very similar to the concrete underlying modeling language. We are convinced that the resulting DSTL is almost as intuitive as the modeling language and provides easy access to novel users. In particular it prevents users from learning the metamodel or abstract syntax of the modeling language.

This systematic process was demonstrated by applying it to the UML/P class diagram language to derive the transformation language CDTrans. The derivation rules defined for this process combine the concrete syntax of class diagrams with a few transformation specific operators. The resulting transformation language is able to describe endogenous model transformations in a domain-specific manner. Although this process was developed based on MontiCore it can also be transferred to other language development frameworks [17]. We assume the presented concept to be helpful to ease the use of transformation languages by untrained developers.